\documentstyle[11pt,epsf,rotate]{article}
\newcommand{\bce}{\begin{center}}
\newcommand{\ece}{\end{center}}
\newcommand{\be}{\begin{equation}}
\newcommand{\ee}{\end{equation}}
\newcommand{\bea}{\begin{eqnarray}}
\newcommand{\eea}{\end{eqnarray}}
\newcommand{\bit}{\begin{itemize}}
\newcommand{\eit}{\end{itemize}}
\newcommand{\E}{\>=\>}
\newcommand{\EA}{&=&}
\newcommand{\deF}{\> =: \>}
\newcommand{\Def}{\> := \>}

\newcommand{\bfl}{\begin{flushright}}
\newcommand{\efl}{\end{flushright}}

\newcommand{\non}{\nonumber \\}

\begin{document}
\thispagestyle{empty}

\vspace*{1cm}

\bce
{\Large \bf Worldline Variational Approximation: A New Approach to the}

\vspace{0.3cm}
{\Large \bf Relativistic Binding Problem}

\vspace{2.5cm}

K. Barro-Bergfl\"odt $^1$, R. Rosenfelder $^2$ and M. Stingl  $^3$

\vspace{0.8cm}
$^1$ Department of Mathematics, ETH Z\"urich, CH-8092 Z\"urich, Switzerland \\
$^2$ Particle Theory Group, Paul Scherrer Institute, CH-5232 Villigen PSI, 
Switzerland \\
$^3$ Institut f. Theoretische Physik, Universit\"at M\"unster,
D-48149 M\"unster, Germany \\

\ece

\vspace{2.5cm}

\begin{abstract}
We determine the lowest bound-state pole of the density-density correlator 
in the scalar Wick-Cutkosky
model where two equal-mass constituents interact via the exchange of mesons. 
This is  done by employing the worldline representation of field theory 
together with a variational approximation as in Feynman's treatment of the 
polaron. Unlike traditional methods based on the Bethe-Salpeter equation, 
self-energy and vertex corrections are (approximately) included as are 
crossed diagrams. Only vacuum-polarization effects of the heavy particles 
are neglected. The well-known instability of the 
model due to self-energy effects leads to large qualitative and quantitative 
changes compared to traditional approaches which neglect them. We determine 
numerically the critical coupling constant above which no real solutions of 
the variational equations exist anymore  and show that it is smaller than in 
the one-body case due to an induced instability. The width of the bound state 
above the critical coupling is estimated analytically.

\end{abstract}

\vspace{1.5cm}

\vspace{0.2cm}


\hspace*{1.7cm} Wick-Cutkosky model

\newpage

\setcounter{equation}{0}

\section{Introduction}
\noindent
Traditionally the relativistic bound-state problem is 
treated in the framework of the Bethe-Salpeter equation \cite{BS} which -- 
although formally exact -- has to be approximated in various ways. 
The most common one is the ladder 
approximation which nearly has become synonymous with {\it the} Bethe-Salpeter 
equation although it has numerous deficiencies. \cite{BSreview}
Over the years three-dimensional reductions, spectator approximations, 
light front methods \cite{overview}-- 
to name just a few variants -- have been investigated and frequently used. 
In hadronic physics where the perturbative methods of bound-state QED 
\cite{QED bs} are of little 
value there is an urgent need for methods which also work at 
strong coupling. Lattice Gauge Theory is considered as the prime method to 
obtain gauge-invariant results from first principles, albeit with enormous 
numerical effort and problems of its own. 
The continous progress of lattice calculations not withstanding,
considerable progress has also been made in the last years in solving 
the Dyson-Schwinger equations for Landau-gauge QCD under some simplifications 
\cite{AlSm} and in describing the low-lying hadrons as 
bound-state of quarks and gluons. \cite{DS} While phenomenologically 
quite succesful and often going beyond the ladder BSE 
these calculations still have limitations due to truncations, 
gauge dependence and the use of model propagators.
Due to that the general impression (at least in the high-energy physics 
community) seems to be that the strong-coupling,
relativistic bound-state problem is just so messy that one has to wait for
better lattice calculations to determine the hadronic masses from the 
binding of quarks and gluons.
Therefore it may be useful to have
a fresh look at this more than $50$-year-old problem from the perspective of 
the particle representation of field theory which has attractive features as 
demonstrated by Feynman's variational treatment of the polaron. \cite{Fey}
Of course, variational methods have also been used before in field theory 
\cite{var} and, in particular for the bound-state problem \cite{Dar}, 
but rather based on fields than on particle trajectories. The
huge reduction in degrees of freedom which the the worldline description 
entails allows to obtain good results with rather crude variational 
{\it ans\"atze}.

\section{Variational worldline approximation to the correlator}

\noindent
``In the relativistic approach, bound states and resonances are identified 
by the occurence of poles in Green functions. A simple extension of the 
Schr\"odinger equation is unfortunately not avaliable $ \ldots$~'' 
(p. 481 in Ref. 10). Therefore 
we look for  poles of a special $4$-point function, the density-density 
correlator (or polarization propagator in the language of many-body theory)
\be
\Pi(q) \Def  - i \, \int d^4x \, e^{i q \cdot x } \, 
\left < 0 \left |\,  {\cal T} \left ( 
\hat \Phi^{\dagger}_2(x) \hat \Phi_1(x)  \,  \hat \Phi^{\dagger}_1(0)  
\hat \Phi_2(0) \right ) \, \right | 0 \right >  
\label{def Pi(q)}
\ee
as a function of the external variable $q$. The correlator (with appropriately
modified currents) is precisely the object from which hadronic masses 
have been estimated in the QCD-sum rule approach. \cite{SVZ} Whereas that method 
uses a delicate matching between short- and long-distance expansions our aim 
is to approximate the correlator variationally and to extract the pole position 
analytically. We do that in the context of the scalar Wick-Cutkosky model 
\cite{WiCu} where heavy charged particles (``nucleons'') 
interact via the exchange of neutral scalar ``mesons'' ($\chi$). 
Its Lagrangian is given by (with $\hbar = c = 1$)
\be
{\cal L} \E \sum_{i=1}^2 \> \left [ \, \left | \partial_{\mu} \Phi_i 
\right |^2 - \left ( M_0^2 + 2 g \chi \right ) \, \left |\Phi_i \right |^2
\, \right ] + \frac{1}{2} \left ( \partial_{\mu} 
\chi \right )^2 - \frac{1}{2} m^2 \chi^2 \> .
\label{L WC}
\ee
Incidentally, this model (with $m = 0$) was just invented to study the 
bound-state problem. In the ladder approximation the Bethe-Salpeter equation 
can be solved exactly (see any field theory textbook which still 
covers bound-state problems). For simplicity, in the present work we assume 
that both particles have the same bare mass $M_0$ but different quantum numbers
so that annihilation into mesons is not possible. The coupling constant 
in Eq. (\ref{L WC}) has been written so as to conform
with previous work in the one-nucleon sector. \cite{WC1}
Bound states of a nucleon of type 1 and an antinucleon of type 2 will
manifest themselves as poles {\it below} the threshold $ q^2 < 4 M^2 $ 
where $M$ is the physical mass of the nucleon. 

\begin{figure}[ht]
\vspace{0.5cm}
\begin{center}
\mbox{\epsfxsize=12cm\epsffile{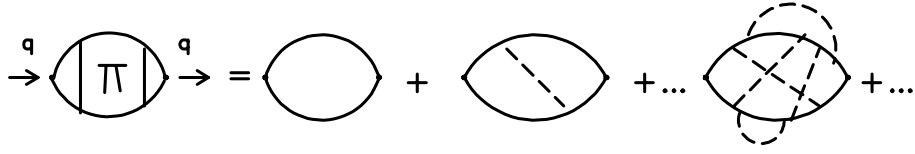}}
\end{center}
\vspace{0.5cm}
\caption{Graphical representation of the correlator $\Pi(q)$ from Eq. 
(\ref{corr}) and its perturbative expansion. Solid lines refer to nucleons, 
dashed ones to mesons. Only diagrams where nucleon pairs are created 
from the vacuum are omitted (quenched approximation).}
\label{fig: correlator}
\vspace{0.5cm}
\end{figure}

In the quenched approximation where nucleon loops are neglected 
(so that there is no divergent nucleon-field and coupling-constant 
renormalization) 
the correlator can be expressed as a double worldline path integral \cite{STG}
\bea
\Pi(q) \EA i \int d^4x \, e^{i q \cdot x} \int_0^{\infty} 
\frac{dT_1  dT_2}{(2 i \kappa_0)^2} \,  
\exp \left [  - 
\frac{i M_0^2 (T_1+T_2)}{2 \kappa_0}   \right ] \non
&& \cdot \> \int  {\cal D} x_1   {\cal D} x_2 \,
\exp \left \{ i \sum_{i=1}^2 S_0[x_i] + 
i S_{\rm int}[x_1,x_2]  \right \} .
\label{corr}
\eea
Here the (4-dimensional) nucleon trajectories have to obey the boundary 
conditions $ x_1(0) = 0 ,$\\  $ x_1(T_1) = x $ and 
$ x_2(0) = x , \, x_2(T_2)= 0 $ ,
\be
S_0[x_i] \E \int_0^{T_i} dt \> \left (  - \frac{\kappa_0}{2} \dot x^2_i(t)
\right ) \> \>, \> \> i \E  1, 2
\ee
is the standard free action for each particle and
\be
S_{\rm int}[x_1,x_2] \E
- \frac{g^2}{2 \kappa_0^2} \, \sum_{i,j=1}^2 \,  
\int_0^{T_i} dt \, \int_0^{T_j} dt' \> \int \frac{d^4 p}{(2 \pi)^4} \> 
\frac{\exp \left [ \, - i p \cdot \left (  x_i(t) -  x_j(t') \right ) \, 
\right ]}{p^2 - m^2 + i0} \> .
\label{Sint 1,2}
\ee
the interaction term. The integration over proper times $T_1, T_2 $ arises 
from the Schwinger representation for each interacting nucleon propagator while
$\kappa_0 > 0$ reparametrizes the proper time and can be considered 
as ``mass'' of the equivalent quantum mechanical particle. The mesons have 
been integrated out exactly which leads to a retarded, two-time action with 
the free meson propagator connecting different points on the worldlines.
If Eq. (\ref{Sint 1,2}) is split into terms with $ i = j $ and $ i \ne j $ one 
sees that the former generate the self-energies of each particle
while the latter describe the interaction between nucleon and antinucleon
by exchange of (any number of) mesons. The vertex corrections come 
automatically due to different values of the proper times; for example, 
if one self-energy meson is already ``in the air'' when another meson 
is exchanged with the second particle (see Fig. \ref{fig: correlator}). 
By the same reason all crossed diagrams are also included.

As in Feynman's treatment of the polaron we approximate the highly nonlinear, 
retarded action by that of a retarded harmonic oscillator with free parameters 
and retardation functions and apply Jensen's inequality 
$ \left < \exp (- \Delta S) \right > \ge \exp ( - \left < \Delta S \right > ) $
(in real time: stationarity, for details see e.g. Ref. 13). 
This is best done in Fourier space where each 
nucleon  trajectory (after redefining $t_2 \to T_2 - t_2$~) can be written as
\be
x_i(t) \E x \, \frac{t}{T_i} + \sum_{k=1}^{\infty} \, 
\frac{\sqrt{2 T_i}}{k \pi} \, 
a_k^{(i)} \, \sin \left ( \frac{k \pi t}{T_i} \right ) \> \> , \> \> i = 1, 2 
\> .
\ee
The functional integration over the trajectories $x_i(t)$ is now replaced
by an integration over the Fourier coefficients $a_k^{(i)}$. 
Including the integral over the final position $x$ into our definition of 
averages we employ the following quadratic trial action
\be
\tilde S_t \E  \tilde \lambda q \cdot x - \frac{\kappa_0}{2} \, \sum_{i=1}^2 \, 
\left [ \, \frac{A_0 x^2}{T_i}  +  
 \sum_{k=1}^{\infty} \, A_k \, a_k^{(i) \> 2} \, \right ] 
+ \kappa_0 \, \sum_{k=1}^{\infty} \, B_k \, a_k^{(1)} \cdot  a_k^{(2)}
\label{tilde St}
\ee
which for $ \> \tilde \lambda = A_k = 1 \> , \> B_k = 0 \> $ reduces to the 
free action. The last term accounts for the direct coupling of the 
two worldlines. Due to the quadratic trial action the 
various averages can be evaluated exactly: 
$ \int \exp (i \tilde S_t) $ and $ \left < \tilde S_0 - \tilde S_t \right > $
give rise to a ``kinetic term'' $ \Omega_{12} $ while 
$ \left < S_{\rm int} \right > $ leads to two distinguished ``potentials''
corresponding to the self-interaction of each particle ( $ V_{ii} $ ) and 
the direct interactions between different particles ( $ V_{ij} , i  \ne j $).
Similar as in previous 
applications of the variational worldline approximation 
a pole develops when the proper time (here the combination $ T = (T_1 + T_2)/2 $
as shown in more detail in Ref. 15)
tends to infinity and only those terms in the exponentials 
contribute which are proportional to $T$.
Then one obtains {\it Mano}'s equation \cite{Mano}
\be
M_0^2 \E 
\left ( \frac{q}{2} \right )^2 \, \left ( 2 \lambda - \lambda^2 \right ) 
- \Omega_{12} - 2 \left( \,  V_{11} + V_{12} \, \right ) \> .
\label{Mano 2}
\ee
Here
$\lambda = \tilde \lambda/A(0) $ is a modified variational parameter. The 
kinetic term $\Omega_{12}$ acts as a restoring term for the variational 
principle and depends on the variational parameters $A_k$ which in the limit 
$T \to \infty$ become a ``profile function'' $A_k \to  A(k \pi/T) \equiv A(E) $. 
Similarly, $B_k \to  B(k \pi/T) \equiv B(E) $.
Finally, the interaction terms
\be 
V_{1j} \E  \frac{g^2}{2 \kappa_0} \, Z^{j-1}
\, \int_{-\infty}^{+\infty} d\sigma\> \int \frac{d^4 p}{(2 \pi)^4} \> 
\frac{1}{p^2 - m^2 + i0} \, \exp \left \{ \, \frac{i}{2 \kappa_0} \, 
\left [ \, p^2 \mu_{1j}^2(\sigma) - \lambda \, p \cdot q \, \sigma \, 
\right ] \, \right \} 
\label{V1j}
\ee
($j = 1,2$) depend on the ``pseudotimes'' $ \mu_{1j}^2(\sigma=t-t') $ 
which basically are 
Fourier cosine transforms of the inverse profile functions. 
To keep track of the binding potential $V_{12}$ we have introduced
an additional factor $Z$ in front of its proper time integral which will 
be set to unity at the end of (analytic) calculations. 
This will allow to distinguish relativistic binding corrections 
from radiative corrections to the binding energy.
A remarkable simplification is achieved by introducing the combinations
$  \> A_{\pm}(E) \Def A(E) \pm B(E) \> $.
Then the kinetic term becomes $ \> \Omega_{12} \E \Omega[A_-] + \Omega[A_+] \> $ 
where 
\be
\Omega[A] \E \frac{2 \kappa_0}{i \pi} \, \int_0^{\infty} dE \> \left [ \, 
\ln A(E) + \frac{1}{A(E)} - 1 \, \right ]
\ee
is just the usual kinetic term encountered in the 
self-energy of a single nucleon \cite{WC1} and the pseudotimes are given by 
\be
\mu_{1j}^2(\sigma) \E \frac{2}{\pi} \, \int_0^{\infty} dE \>  \frac{1}{E^2} \, 
\left \{ \, \frac{\delta_{j2}}{A_+(E)} + \left [ \, 
\frac{1}{A_-(E)} + \frac{(-)^{j+1}}{A_+(E)} \, \right ] \, \sin^2 \left ( 
\frac{E \sigma}{2} \right ) \, \right \} \> .
\label{mu1j by Aplusminus}
\ee
Closer inspection of the variational equations (see below) reveals that 
for small proper times $\mu_{11}^2(\sigma) \to \sigma $ 
exhibits the usual self-energy behaviour 
but that $\mu_{12}^2(\sigma) \to {\rm constant}$ which is a new feature 
due to binding. This implies that only the self-energy part $V_{11}$ 
develops divergencies which after regularization with a cutoff $\Lambda$ 
can be absorbed into a mass renormalization
\be
M_0^2 \> \longrightarrow \> M_1^2 \E M_0^2 - \frac{ g^2}{4 \pi^2} 
\, \ln \left ( \frac{\Lambda^2}{m^2} \right ) \> ,
\ee
{\it exactly} as in the one-body case. Note that the intermediate mass $M_1$ is 
not yet the physical mass $M$ which is obtained from Mano's one-body 
equation by a finite shift.

\vspace{0.2cm}

\section{Numerical results and induced instability}

\begin{figure}[ht]
\vspace{0.5cm}
\begin{center}
\mbox{\epsfysize=9.5cm\epsffile{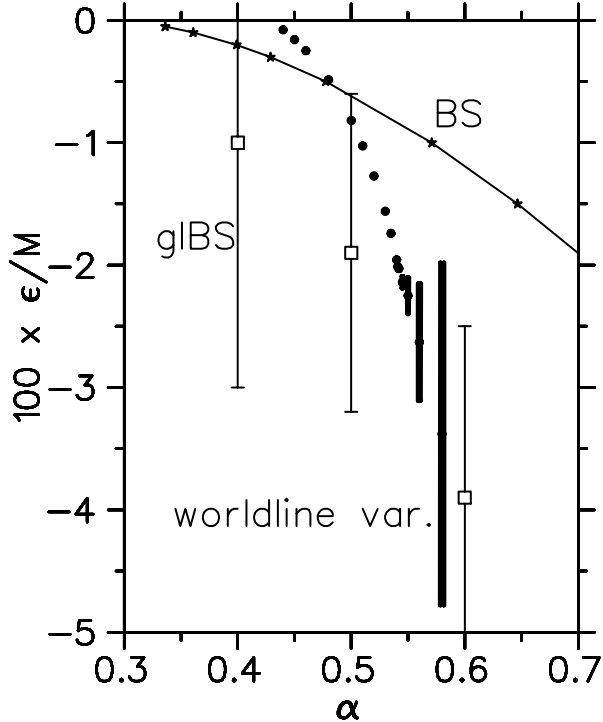}}
\end{center}
\vspace{0.5cm}
\caption{The binding energy  $\epsilon/M $ of the two-body bound state
for $ M = 0.939 $ GeV, $ m = 0.14 $ GeV as a function of the dimensionless
coupling constant defined in Eq. (\protect\ref{def alpha}).
The results from the worldline variational approach including self-energy and 
vertex corrections are compared with those from 
Efimov's variational approximation to the ladder 
Bethe-Salpeter (BS) equation. \protect\cite{Efi} Also shown are the Monte-Carlo 
results (with errors)  for $m/M = 0.15$ from Ref. 19 in 
the ``generalized ladder approximation'' to the 
Bethe-Salpeter equation (glBS). The thick bars for the worldline 
results represent the width of the bound state above the critical coupling 
estimated from Eq. (\protect\ref{Gamma near crit}).}

\label{fig: binding}
\vspace{0.3cm}
\end{figure}

\noindent
By construction Mano's Eq. (\ref{Mano 2}) is stationary under variation of 
variational parameters and functions. Performing the variation w.r.t. $\lambda$
one obtains
\be
\lambda \E 1 - \frac{4}{q^2} \, \frac{\partial}{\partial \lambda} \left ( 
V_{11} + V_{12} \right) \> ,
\label{var eq for lambda 1}
\ee
and variation w.r.t. the profile functions $A_{\pm}(E)$ gives
\bea
A_-(E) \EA 1 + \frac{2i}{\kappa_0} \, \frac{1}{E^2} \, \int_0^{\infty} d\sigma
\>  \sum_{j = 1}^2 \, \frac{\delta V_{1j}}{\delta \mu^2_{1j}(\sigma)} 
\, \sin^2 \left ( \frac{E \sigma}{2} \right ) 
\label{var eq for Aminus}\\
A_+(E) \EA 1 + \frac{2i}{\kappa_0} \, \frac{1}{E^2} \, \int_0^{\infty} d\sigma
\>  \left [ \, \frac{\delta V_{11}}{\delta \mu^2_{11}(\sigma)} 
\, \sin^2 \left ( \frac{E \sigma}{2} \right )  + 
\frac{\delta V_{12}}{\delta \mu^2_{12}(\sigma)} 
\, \cos^2 \left ( \frac{E \sigma}{2} \right ) \, \right ]  .
\label{var eq for Aplus}
\eea
The derivatives of the interaction terms are easily worked out and are not given 
here. Note that the pseudotimes are related to the profile functions by Eq. 
(\ref{mu1j by Aplusminus}).
We have solved these coupled variational equations 
for $\lambda, A_{\pm}(E)$ and 
$\mu_{1j}^2(\sigma)$ numerically in euclidean time \footnote{This can be
simply obtained by setting $ \kappa_0 = i $ and reversing the sign of 
all four-vector products.}
with similar iterative methods as described in Ref. 17.
Indeed, we have found a pole at $q_0^2 < 4 M^2 $ for certain ranges of the 
mass ratio $m/M$ and the standard dimensionless coupling constant
\be
\alpha \E \frac{g^2}{4 \pi M^2} \> .
\label{def alpha}
\ee
While more details will be given elsewhere  \cite{BRS}, some results for the 
binding energy $ \epsilon = \sqrt{q_0^2} - 2 M $ are shown in Fig. 
\ref{fig: binding}. The values of the intermediate mass $M_1$ have been taken 
from Table III in Ref. 17.

It should be remembered that in the non-relativistic quantum mechanics of 
two particles interacting via an attractive Yukawa potential
(to which our model reduces in the limit $c \to \infty$) binding only occurs 
for $ \delta := \alpha \, M/m  >  1.67981 $. \cite{Pol} The variational 
approximation with a harmonic oscillator potential requires 
$ \delta > 2.7714 $ since it provides only an upper limit for the 
binding energy. Thus for a finite pion mass a minimal coupling strength 
is needed to obtain a bound state. Beyond this threshold value we observe 
much stronger binding when self-energy, vertex corrections and crossed 
diagrams are taken into account.
At first sight this is similar to the results obtained 
in the ``generalized ladder approximation'' to the 
Bethe-Salpeter equation~\cite{NiTj} in which all ladder and crossed-ladder 
diagrams have been included. However, in these Monte-Carlo calculations 
self-energy and vertex corrections are still 
neglected which allows binding for arbitrary large coupling constants. 

Self-energy corrections to the light-cone Tamm-Dancoff approximation
for two nucleons have been considered by Ji \cite{Ji} 
who found these corrections to act in a repulsive way.
Although approximations in the light-cone and the equal-time formalism
are difficult to compare it seems to us that this statement is not valid. 
This can be best seen in the weak-coupling case for massless mesons where the 
binding energy of two equal-mass nucleons takes the form
\be
\frac{\epsilon}{M/2} \E - b_2 \, ( Z \alpha )^2 \, \left [ \, 1 + r_{21} \, 
\frac{\alpha}{\pi} + \ldots \> \right ] - b_4 \, ( Z \alpha )^4 \, \Bigl  [ \, 
1 + \ldots \, \Bigr ] - \ldots \> \> .
\label{weak coup}
\ee
The exact values of the binding coefficients are $b_2 = 1/2$ (from the 
non-relativistic Coulomb problem) and $b_4 = 5/32$ (from Todorov's equation for 
``scalar photons''~\cite{BIZ}). 

As we include (approximately) self-energy and vertex corrections in our 
bound-state calculation it is worthwhile to discuss the radiative coefficient 
$ r_{21} $ in more detail: due to vertex corrections the effective coupling 
constant is enhanced, i.e. $ r_{21} $ is positive and {\it increases} the 
binding. This enhancement was obtained in Eq. (58) of Ref. 23 and 
is also the exact one-loop result
for a free nucleon because the first-order variational calculation 
reduces to that in the weak-coupling limit. Alternatively, by standard 
Feynman-diagram techniques one may calculate
the physical amplitude for meson-nucleon scattering at $ q = 0 $ 
\be
T(q \to 0) \E Z_r \, \Gamma(p,q \to 0) \Bigr |_{p^2 = M^2}  \deF 
g_{\rm eff} 
\label{rel amplitude}
\ee
where $Z_r$ is the residue of the 2-point function at the pole and 
$ \Gamma(p,q) $ the truncated meson-nucleon amplitude. From the one-loop 
diagrams for these quantities one easily obtains
\be
g_{\rm eff} \E 
g \, \left [ \, 1 + \frac{g^2}{4 \pi^2} \, \int_0^1 dx \> \frac{x^2}{ M^2 x^2 + 
m^2(1-x)} \right ]  \> \stackrel{m = 0}{\longrightarrow} \> 
g \, \cdot  \, \left ( \, 1 + \frac{\alpha}{\pi} \right ) \> .
\label{g eff}
\ee 
From Eq. (\ref{def alpha}) we therefore deduce that 
$ \alpha \to \alpha ( 1 + 2 \alpha/\pi + \ldots ) $ and that 
the radiative correction to the coulombic $(Z \alpha)^2$-term is given 
by $ ( 1 + 4 \alpha/\pi + \ldots ) $, i.e. $r_{21} = 4 $.
Note that this procedure corresponds to the one-loop determination 
of the Wilson-coefficient $c_1$ 
in the effective non-relativistic field theory of Ref. 24.
In such a description explicit antiparticle degrees of freedom and 
high-energy modes are ``integrated out'' 
but their effect is retained in the coefficients of the effective theory. 
Matching the non-relativistic meson-nucleon scattering amplitude with the
one-loop relativistic amplitude for nucleon three-momentum $ {\bf p} \to 0 $ and 
three-momentum transfer $ {\bf q} \to 0 $ indeed determines
$c_1 = g_{\rm eff}/g $. 

What does the worldline variational method predict in the weak-coupling limit
for massless mesons~? The answer can be obtained by solving the variational 
equations (\ref{var eq for lambda 1}) - (\ref{var eq for Aplus})
analytically in that limit. As shown elsewhere \cite{BRS}
this gives $ \> b_2^{\rm var} = 1/\pi \> , \> 
r_{21}^{\rm var} = 7/2 \> , \>  b_4^{\rm var} = 1/\pi^2 \> $.

In view of the fact that both effective field theory and the present approach 
predict {\it more} binding (although the latter one  with smaller numerical 
coefficients as expected from a variational calculation)
serious doubts remain whether Ji's
light-cone calculation is correct. It is unclear to us if
this is due to an incomplete nucleon mass renormalization (for a 
recent discussion  
see Ref. 25) or because of the restrictive Tamm-Dancoff approximation.

\vspace{0.3cm}

In the worldline variational method an upper limit for the coupling constant 
comes from the well-known instability of the Wick-Cutkosky model. 
\cite{Baym,unstable} 
In the one-body case (i.e. for the two-point function) the 
polaron variational approximation failed to have real solutions for coupling 
strengths beyond \footnote{Similar values have been obtained in truncated
Dyson-Schwinger calculations. \cite{AhAl} Below the critical coupling 
the exact theory would still generate a nonzero, but presumably exponentially 
small decay width for the particle. Readers uncomfortable 
with the notion of an unstable ground state (asking ``into what does it 
decay ?'')
may replace the phrase ``decay width'' by ``inverse lifetime of the metastable 
state prepared at $t = 0$ ``. For recent discussions of the instability
see Refs. 27 and 29.} 
\be
\alpha_{\rm crit}^{(1)} \E 0.815 \> .
\label{alpha crit 1}
\ee

\noindent
Although a quadratic trial action is incapable of describing 
tunneling phenomena \cite{AlSa} 
this is a genuine nonperturbative result 
which cannot be obtained by an expansion in powers of $\alpha$.
In the present calculations we have found the same phenomenon 
at {\it smaller} values of $\alpha$
\be
\alpha_{\rm crit}^{(2)} \simeq 0.54 
\label{alpha crit 2}
\ee
implying an {\it induced} (or catalyzed) instability \cite{induced} due to the 
presence of an additional particle. This can be understood by a particular 
simple variational {\it ansatz} first employed in the one-body case \cite{WC1}: 
$A_-(E) = 1 \> , \> \lambda $ free and, additionally 
\be
A_+(E) \E 1 + \frac{\omega^2}{E^2} \> .
\label{ansatz A+}
\ee
Eq. (\ref{ansatz A+}) is exactly the euclidean profile function 
for an harmonic oscillator
and accounts for the binding of the particles. Assuming weak binding 
($ q^2 \simeq 4 M^2, \omega \ll M^2 $) and massless mesons ($m = 0 $) 
the variational equations for the parameters $\omega, \lambda $ 
become simple algebraic equations. In particular, the one for 
the parameter $\lambda $ is a quartic equation
\be
\lambda^4 - \lambda^3 + \frac{\alpha}{2 \pi} \, \lambda^2 + 
\frac{(Z \alpha)^2}{2 \pi} \E 0 
\label{quartic eq}
\ee
which generalizes the quadratic equation for the one-body case. Indeed, by 
setting $ Z = 0 $ one obtains the former estimate 
$ \alpha_{\rm crit}^{(1)} \simeq  \pi/4 = 0.785$. It is a simple exercise to
determine the critical coupling constant from Eq. (\ref{quartic eq}) and one 
finds
\be
\alpha_{\rm crit}(Z) \> \simeq \>  \frac{\pi}{8} \, 
\frac{(1 + \sqrt{1 + 3 z})^3}{(1 + z + \sqrt{1 + 3 z})^2} \> \le \, 
\frac{\pi}{4}  \> \> , \hspace{0.5cm} z \Def 2 \pi Z^2 
\label{alpha crit}
\ee
which for $Z = 1$ gives $ \alpha_{\rm crit}^{(2)} \simeq 0.463$ in fair 
agreement with the numerical result (\ref{alpha crit 2}).

It is also possible to determine approximately the width 
$ \Gamma$ of the bound state for $ \alpha > \alpha_{\rm crit}^{(2)} $ 
following the very same treatment as in section IV. A of Ref. 17. 
The idea is to look for a {\it complex} analytical solution of the variational 
equations with the simplified ansatz (\ref{ansatz A+}). This is only possible
if the physical mass acquires an imaginary part, i.e. a width.
Close to the critical coupling constant one then finds
\be
\Gamma  \> \simeq \>  \frac{2}{3} \, 2 M \, 
 \left ( \frac{\alpha - \alpha_{\rm crit}^{(2)}}{\alpha_{\rm crit}^{(2)}} 
\right )^{3/2} \, \cdot \, f_{\rm corr} (Z)
\label{Gamma near crit}
\ee
where the correction factor $f_{\rm corr}$ varies between $1$ for $ Z = 0 $ and 
$1.0885$ for $ Z = 1$.
In view of the rough approximations employed to obtain this result the value of 
the correction factor is of much less significance than the mass factor $2 M$ in 
Eq. (\ref{Gamma near crit}) since in the one-body case it simply was $M$. 
Thus the instability induced by the presence of the second particle 
not only shows up in the lower critical coupling constant but also in a much 
larger width above it.

\newpage
\section{Summary and outlook}

\noindent
We have extended the worldline variational method to the relativistic binding 
problem of two equal-mass particles in the scalar Wick-Cutkosky model and 
obtained binding energies which include self-energy, vertex and retardation 
effects consistently. In this way not only increased binding due to 
radiative corrections 
was obtained in the weak-coupling case but also the physics
of strong coupling could be addressed. In the latter case, 
an enhanced instability due to the presence of 
the second particle was found as a non-perturbative effect in this 
(admittedly unrealistic) model.
From a quantitative viewpoint the major drawback of the present worldline 
variational method (apart from the
quenched approximation) is the relatively poor description of the 
binding interaction by means of a quadratic trial action. Again, this can be 
best seen in the weak-coupling expansion (\ref{weak coup})
where for $ m = 0$ not the exact Coulomb result for the leading term is 
obtained. 
Different methods developed for the polaron problem could be applied to improve 
on that: a more general quadratic trial action \cite{WC7},
second-order corrections to Feynman's result \cite{MaMi}, Luttinger \& Lu's 
improved variational {\it ansatz} \cite{LuLu} or Monte-Carlo simulations 
\cite{NiTj,MC} are possible directions for further work. 
Extension to gauge theories looks promising since in the one-body sector 
the worldline variational method does not depend on the covariant gauge 
parameter \cite{QED var} in contrast to the Dyson-Schwinger approach. \cite{BaRa}
However, already in the present formulation the variational worldline 
approximation gives novel results for the relativistic bound-state problem 
by treating in one consistent approximation the two-point (self-energy), 
three-point (vertex correction), and four-point (binding) effects.

\end{document}